# Anomalous Translational Dynamics of Molecular Probes Near the Polymer Glass Transition

Jaladhar Mahato, Siyang Wang, Laura J. Kaufman[*]

Department of Chemistry, Columbia University, New York, NY 10027

*[*]To whom correspondence should be addressed: kaufman@chem.columbia.edu

## Abstract

The origin of the dramatic slowdown of dynamics near the glass transition temperature ($T_g$) remains a long-standing fundamental and unresolved issue in soft condensed matter. While single-molecule (SM) experiments using fluorescent probes have provided critical insight for molecular and polymeric glass formers through rotational measurements, translational dynamics remain largely unexplored in such systems at the molecular length scale. Here, we report SM translational dynamics of molecular probes in high molecular weight polystyrene at three temperatures near $T_g$. The probes exhibit quasi-stationary position fluctuations, non-Gaussian displacement distributions, sub-diffusive transport with anti-correlated displacements, and a characteristic translational relaxation time. The observations are quantitatively described using a microscopic framework based on the generalized Langevin equation and supported by numerical modeling for heterogeneous transport. The probe's translational dynamics provides direct microscopic evidence of dynamic heterogeneity and suggests a pathway to more fully understand glassy dynamics in glass formers near $T_g$.

## Introduction

Disordered materials formed by rapid cooling of liquids below their melting temperature are of great interest owing to their wide technological application and unusual fundamental properties. While many liquids require careful thermal control to bypass crystallization, polymers and many organic liquids can readily form glasses (1, 2). It is well established that below the glass transition temperature ($T_g$) glass formers falls out of equilibrium (3); however, near and above $T_g$, such systems are generally considered to exist at local thermal equilibrium within the metastable supercooled state (2, 4). In the supercooled regime, such disordered systems exhibit unusual behaviors, including (i) dramatic change in viscosity upon small changes in temperature, reflecting non-Arrhenius relaxation; (ii) stretched exponential relaxation dynamics suggesting a broad distribution of relaxation timescales; and (iii) apparent rotational-translational decoupling as the system approaches $T_g$ (2, 5). Both ensemble experiments and computer simulations across glass formers consistently show these features near $T_g$. The distribution of relaxation timescales, i.e. dynamic heterogeneity, in which subsets of molecules with distinct relaxation dynamics are distributed over space and time, complicates the formulation of predictive theories (1, 5, 6).

Single-molecule (SM) experiments in which probes report the behavior of the surrounding supercooled liquid provide microscopic dynamical information by analyzing dynamics of particular regions over time, potentially revealing heterogeneities hidden in ensemble-averaged approaches. SM methods are particularly powerful for systems like supercooled liquids and glasses where no obvious structural inhomogeneity is present, as they enable direct investigation of dynamic heterogeneity and its role in glassy dynamics near $T_g$ (7, 8). With appropriate probe selection, SM measurements successfully demonstrated non-Arrhenius temperature dependence of rotational dynamics of these systems while simultaneously revealing significant heterogeneity evident from the broad distribution of probe rotational relaxation times, $\tau_c$, and sub-unity stretching exponents of SM autocorrelation functions (9, 10). Moreover, one of the few experiments probing translational dynamics in a supercooled liquid showed that the temperature dependence of rotational and translational dynamics deviates noticeably from Stokes−Einstein (SE) and Debye−Stokes−Einstein (DSE) predictions even at the SM level (11).

While SM translational measurements have rarely been used to study glassy materials, position fluctuation of individual probes can provide important insights into such materials (12–18). In heterogeneous systems, the transport is often anomalous: displacement distributions deviate from Gaussian statistics (12, 13, 17–21), and the mean-squared displacements often display signatures of sub-diffusion (12, 15, 17–24). In glass-forming materials, at the single particle level translational dynamics have primarily been explored in systems with relatively large particles such as colloidal glasses and granular materials (20, 21). These systems are experimentally advantageous because the large particle sizes allow direct visualization of particle rearrangements. In contrast, the length scales of interest in molecular systems near $T_g$ are below the optical diffraction limit, complicating imaging and analysis. However, SM localization microscopy of fluorescent probes in systems near $T_g$ does allow localization with an accuracy of a few nanometers (25). Indeed, a previous SM study was successfully performed and revealed dramatic deviation from standard transport laws: in high molecular weight polystyrene, the translational diffusion of molecular probes was found to be enhanced by approximately 400-fold relative to SE/DSE predictions (11). Here, we analyze similar data to directly investigate the microscopic mechanism of translational transport in polystyrene near $T_g$.

In particular, we report SM translational dynamics of the molecular fluorescent probe: N,N'-dipentyl-3,4,9,10-perylenedicarboximide (pPDI) in high molecular weight polystyrene near $T_g$ (374.3K). We present evidence that probes exhibit position fluctuations consistent with the local equilibrium picture near $T_g$. As the system approaches $T_g$, displacement distributions become increasingly non-Gaussian and show evidence of spatial as well as time-varying heterogeneous dynamics. Our detailed analysis further reveals anti-correlated displacements with confined sub-diffusive behavior. We provide a microscopic framework based on the generalized Langevin equation and a numerical model consistent with these experimental observations. We envision that these findings on SM translational dynamics will offer a new perspective for understanding the transport and relaxation mechanisms in glassy systems, while complementing existing studies.

**Experimental Results**

**Time-reversibility:** Fig. 1a shows representative SM trajectories $x(t)$ recorded at three different temperatures near $T_g$: 380.6K, 377.6K, and 374.5K (see Materials and Methods). While the fluctuation magnitude varies among individual trajectories, the trajectories typically remain bounded, suggesting that probes are spatially confined on the timescale of the experiment. Following earlier reports, we tested the time-reversibility of the data by analyzing higher-order correlation functions from sufficiently long $x(t)$ at all temperatures (22, 23). Specifically, we computed the fourth-order moments, $\langle x^3(0)x(\tau)\rangle$ and $\langle x(0)x^3(\tau)\rangle$. Fig. 1b shows that their scatter plots lie along the diagonal, which indicates $\langle x^3(0)x(\tau)\rangle = \langle x(0)x^3(\tau)\rangle$, a hallmark of time-reversal symmetry and the stationarity of the $x(t)$ process. The observation is consistent across all temperatures. Furthermore, we computed the third-order moment, at two different intermediate points, $\langle x(0)x(\tau)x(3\tau)\rangle$ and $\langle x(0)x(2\tau)x(3\tau)\rangle$, shown as a scatter plot in Fig 1c. For a time-reversible process, these two quantities should be equal, and for a Gaussian process, they should vanish when averaged over many trajectories, i.e., $\overline{\langle x(0)x(\tau_1)x(\tau_2)\rangle} \approx 0$. Fig. 1c shows such results again fall along the diagonal, proving the time-reversibility of $x(t)$. While ensemble symmetry is preserved, as evident from the negligible third-order correlations, the growing spread of data from individual trajectories as the temperature approaches $T_g$ suggests that, within our experimental timescale, the individual probes may not relax sufficiently to recover Gaussian behavior.

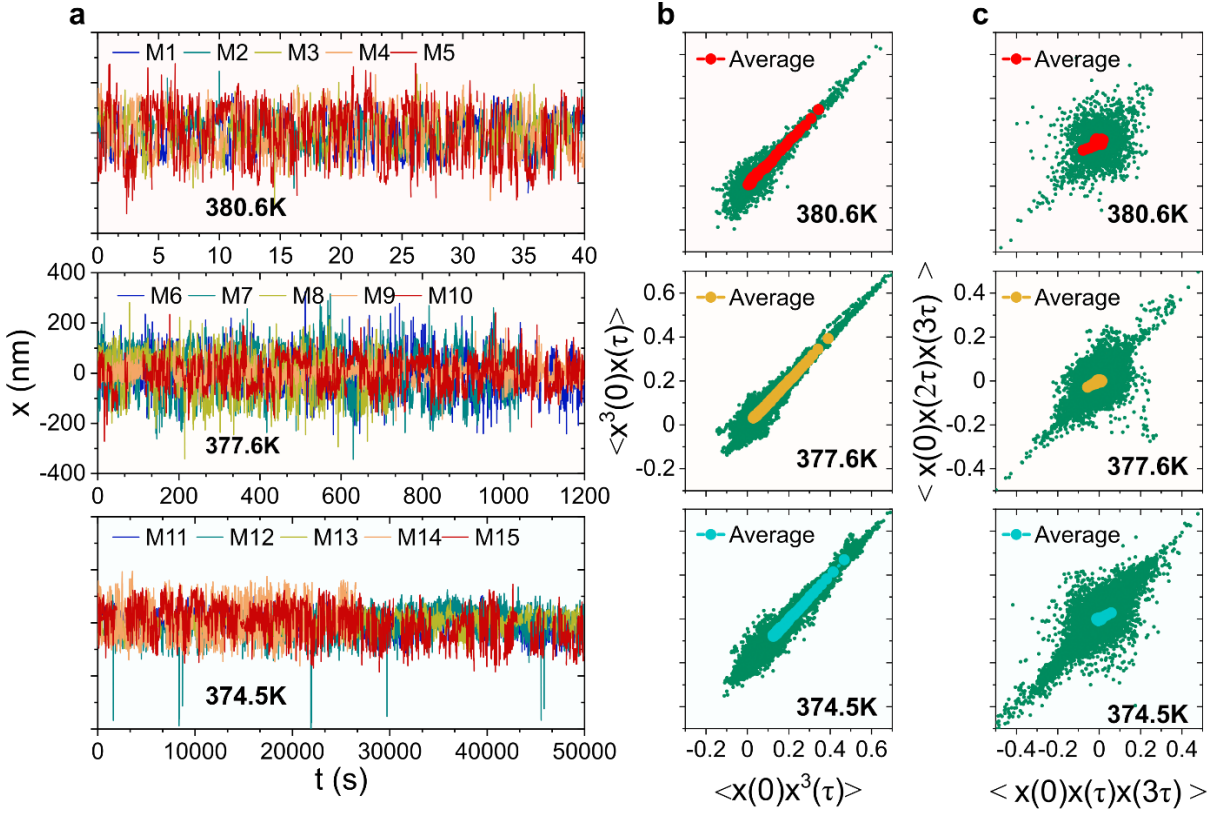

**Fig. 1.** Translational dynamics of single-molecule (SM) probes in high molecular weight polystyrene at three temperatures near $T_g$: 380.6K, 377.6K, and 374.5K. (a) Five representative mean-subtracted SM trajectories illustrating diverse dynamics. (b, c) The normalized higher order correlations. Scatter plots of (b) $\langle x^3(0)x(\tau)\rangle$ vs $\langle x(0)x^3(\tau)\rangle$ to assess time-reversibility and (c) $\langle x(0)x(\tau)x(3\tau)\rangle$ vs $\langle x(0)x(2\tau)x(3\tau)\rangle$ to assess both time-reversibility and relaxation toward Gaussian behavior. All higher order correlations were computed from more than 50 individual trajectories (dark green) and their averages are shown as solid lines: red for 380.6K, yellow for 377.6K, and cyan for 374.5K.

**Displacement distribution:** To further study translational motion in these systems, we examined the distribution of successive displacements $\Delta x(\varepsilon = m\Delta t) = x(t + \varepsilon) - x(t)$, where $m = 1, 2, 3, \ldots N - m$ with $N$ denoting the total number of time points in each SM trajectory. Fig 2a shows displacement distributions, $P(\Delta x, \varepsilon)$, compiled from all available SM trajectories. The spread of $P(\Delta x, \varepsilon)$ decreases upon approaching $T_g$. At all temperatures, the distributions are clearly non-Gaussian. While smaller displacements are prominent at lower temperature, broad non-Gaussian tails are observed at all conditions, indicative of heterogeneous dynamics (13, 19). To understand the effect of probe-specific local viscosity, we normalized each mean-subtracted trajectory by its own standard deviation and recomputed the displacement distribution, $P(\Delta x/\sigma, \varepsilon)$. If the observed non-Gaussianity arises due to intrinsic variation in the friction experienced by different probes, such normalization can yield Gaussian statistics. As shown in Fig. 2b, this normalization substantially reduces the contribution near the origin and at short time lags, yet the dynamics remain non-Gaussian, especially as $T_g$ is approached. Interestingly, at longer time lags, near $\varepsilon \approx 20\tau_c$, with $\tau_c$ the rotational relaxation time of pPDI in polystyrene (11), $P(\Delta x/\sigma, \varepsilon)$ becomes more Gaussian, as shown by the solid red line in Fig. 2b. As such long lag times correspond to the environmental exchange timescale of the system (9), the obtained results indicate that inhomogeneities average out over longer timescales to restore Gaussian statistics (13, 19). To further understand this temporal variation, we segmented the long SM trajectories based on moving variance (see Methods). Subsequently, each segment was normalized by its own standard deviation before constructing the displacement distribution, $P(\Delta x/\sigma_s, \varepsilon)$. As shown in Fig. 2c, this procedure yields approximately Gaussian $P(\Delta x/\sigma_s, \varepsilon)$, indicating a locally linear dissipation mechanism.

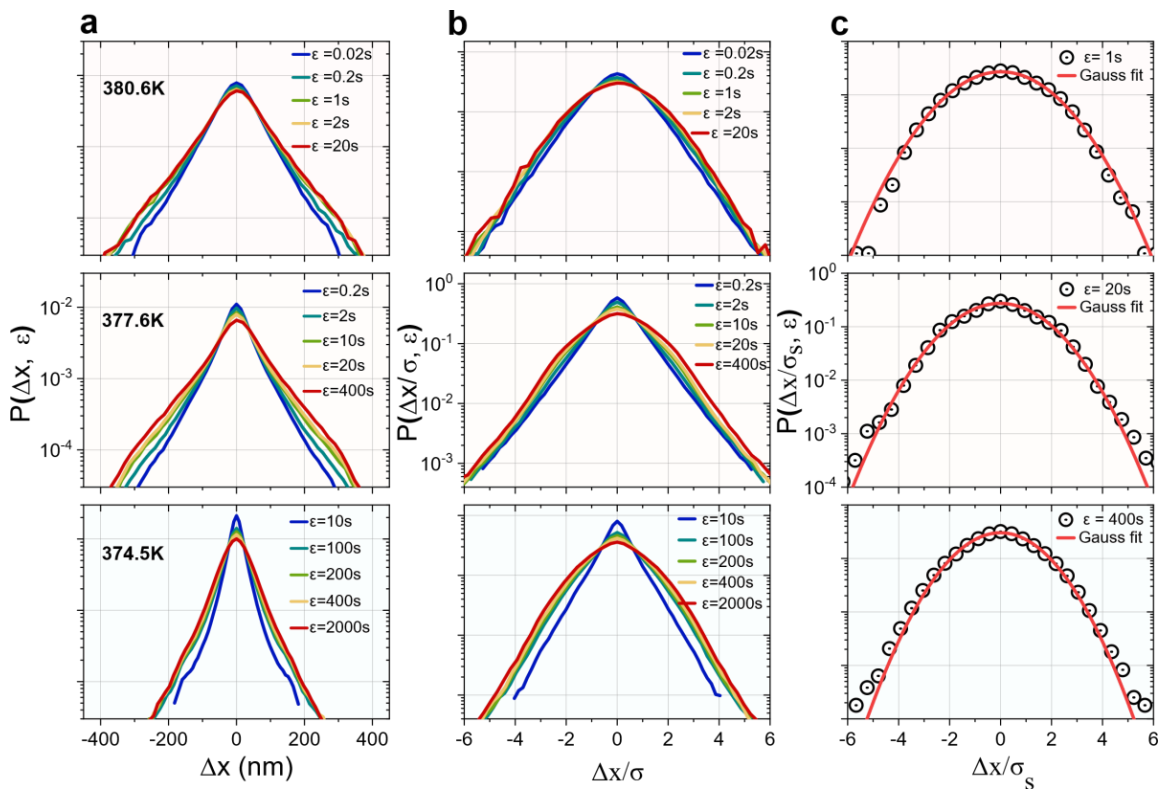

Fig. 2. Displacement distributions of SM probes, defined as $\Delta x(\varepsilon = m\Delta t) = x(t + \varepsilon) - x(t)$ at three temperatures near $T_g$. (a) Probability density functions, $P(\Delta x, \varepsilon)$, of displacements. (b) Probability density functions, $P(\Delta x/\sigma, \varepsilon)$, constructed from the trajectories that are mean-subtracted and normalized by their respective standard deviations $\sigma$ of $x(t)$. (c) Probability density functions, $P(\Delta x/\sigma_s, \varepsilon)$, constructed following a procedure in which the trajectories are segmented as described in Methods, and subsequently each segment is mean-subtracted and normalized by the standard deviation within that window. We observed that $m > 50$ time points, corresponding to 1-10 $\tau_c$, are needed at our experimental frame rate to recover Gaussian $P(\Delta x/\sigma_s, \varepsilon)$. Here, $\varepsilon$ represents the time-window size.

**Mean-squared displacement:** To quantify the transport behavior, we evaluated the time-averaged mean-squared displacement (TAMSD) of individual trajectories defined as (14–18):

$$\langle \Delta x^2(\varepsilon = m\Delta t)\rangle = \frac{1}{N-m} \sum_{t=1}^{N-m} [x(t+\varepsilon) - x(t)]^2 \quad (1)$$

To avoid statistical bias at longer lag times, only probes with sufficiently long trajectories ($m \geq 50$) were analyzed, and their TAMSDs are shown in Fig. 3a. We find that the TAMSDs span nearly a decade in amplitude, implying substantial variation of local environments sampled by the individual probes – a feature consistently observed across all the temperatures. Additionally, the vast majority of probes exhibit anomalous TAMSD scaling $\langle \Delta x^2(\varepsilon) \rangle \propto \varepsilon^\alpha$ with $\alpha < 1$ at short lag times, indicating sub-diffusive transport. Moreover, the average of such TAMSDs shows downward curvature at longer lag times even in the log-log representation, indicative of confinement on our experimental timescales. Altogether, the TAMSD analysis suggests individual probes experience sub-diffusive and possibly confined motion in highly locally heterogeneous environments.

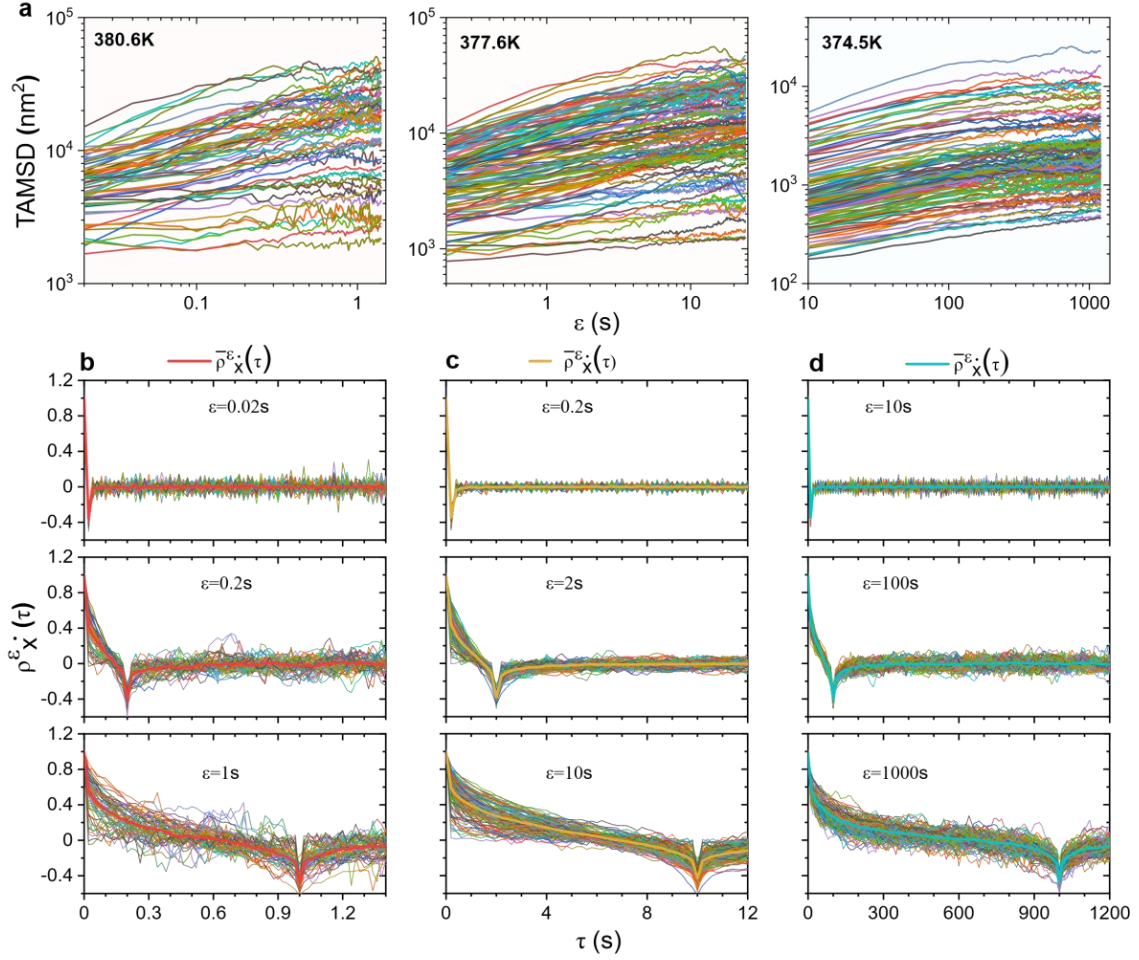

Fig. 3. (a) Time-averaged mean-squared displacements (TAMSDs), $\langle |\Delta x^2(\varepsilon = m\Delta t)| \rangle$ of SM probes at three temperatures near $T_g$: 380.6K, 377.6K, and 374.5K. We computed TAMSDs from trajectories with $m \geq 50$ and retained those TAMSDs that exhibited an initial positive slope ($\alpha > 0$) on the log-log scale and a high coefficient of determination ($R^2 > 0.9$) when fit. Probe transport shows sub-diffusive behavior ($\alpha < 1$) and confinement. (b–d) Normalized displacement rate correlation functions, $\rho_{\dot{x}}^{(\varepsilon)}(\tau)$, of SM probes (thin lines) and their averaged curves (thick lines),

where $\tau = n\Delta t$ and $\dot{x} = \Delta x(\varepsilon)/\varepsilon$ are evaluated near $T_g$. Results are shown for three $\varepsilon$ values. (b) T = 380.6K at $\varepsilon = 0.02s, 0.2s, 1s$. (c) T = 377.6K at $\varepsilon = 0.2s, 2s, 10s$. (d) T = 374.5K at $\varepsilon = 10s, 100s, 1000s$.

**Displacement rate correlation:** Correlation between successive displacement steps, $\Delta x(\varepsilon)$, provides valuable insight into the nature of underlying transport processes. To quantify such potential correlation, we computed the normalized autocorrelation functions of the displacement rate $\Delta x(\varepsilon)/\varepsilon$, analogous to velocity correlation functions:

$$\rho_{\dot{x}}^{(\varepsilon)}(\tau = n\Delta t) = \frac{\langle \Delta x^\varepsilon(t+\tau)\Delta x^\varepsilon(t)\rangle}{\langle [\Delta x^\varepsilon(t)]^2\rangle}. \tag{2}$$

This quantity was calculated from the displacement trajectories at a fixed window $\varepsilon = m\Delta t$. As shown in Fig. 3b-d, most SMs show a pronounced negative peak of $\rho_{\dot{x}}^{(\varepsilon)}(\tau)$ at $m = 1$, which indicates anti-correlated adjacent displacements (15, 24, 26). This anti-correlation is consistently observed for all measured temperatures across trajectories. Such behavior is in stark contrast to ordinary Brownian diffusion, where successive displacement steps are uncorrelated and the autocorrelation decays to zero without a negative dip. Remarkably, the negative correlation persists with increasing $\varepsilon$ and reaches its most negative peak when $\tau = \varepsilon$. The persistence of the negative correlation after averaging over many trajectories, $\overline{\rho_{\dot{x}}^{(\varepsilon)}(\tau)}$, and its consistency across $\varepsilon$ underscores its robustness and suggests the limited influence of measurement error. Furthermore, the gradual increase in the amplitude of the negative correlation peak with increasing $\varepsilon$ suggests that the probes experience a restoring force consistent with confinement (27). Taken together, the displacement rate correlation analysis implies temporally correlated dynamics and probe confinement.

**Transport statistics and challenges:** The apparent confinement implies that the probe dynamics can be characterized by a relaxation timescale, $\tau_D$, and a set of sub-diffusion parameters $\{D_\alpha, \alpha\}$, where $D_\alpha$ is the generalized diffusion coefficient. In principle, $\tau_D$ and $\alpha$ can be estimated from a sufficiently long-time TAMSD curve, while $D_\alpha$ and $\alpha$ can be obtained from the earlier time-lag regime where the probe's motion is minimally influenced by a confining potential. In the short-time regime, $D_\alpha$ and $\alpha$ can be determined from linear fits to the logarithm of SM TAMSDs with $\alpha > 0$ (see Materials and Methods). As shown in Fig. S1 (SI Appendix), $D_\alpha$ shows a log-normal distribution and $\alpha$ a normal distribution across all temperatures, consistent with our previous work, although here the reported generalized diffusion coefficients are somewhat higher owing to differences in data analysis (see Materials and Methods) (11). The resulting median values of $logD_\alpha$ are 3.94 nm²s⁻ᵅ, 3.45 nm²s⁻ᵅ, and 2.18 nm²s⁻ᵅ with corresponding full widths at half maximum (FWHM) of 0.82, 0.80, and 0.92 at 380.6K, 377.6K and 374.5K, respectively. The standard deviation of the $\alpha$ distributions are $\sigma_\alpha \approx 0.1$ across all temperatures, with $\langle\alpha\rangle =$ 0.31, 0.28, and 0.28 at 380.6K, 377.6K and 374.5K, respectively.

Notably, previous studies have shown that early-time TAMSD values are affected by noise-limited probe localization and hence can bias $\{D_\alpha, \alpha\}$ statistics (27, 28). In our experiments, we maintained high signal to noise ratio (S/N) to ensure localization error of $\approx$10 nm, thereby minimizing such artifacts (11). Additionally, SM trajectories produced noisy TAMSDs at long time lags, resulting in unreliable estimation of $\tau_D$. To mitigate TAMSD related limitations, we analyze $\rho_{\dot{x}}^{(\varepsilon)}(\tau)$, which present an entire function (rather than a single point) at a specific $\varepsilon$. The parameters $\{\tau_D, \alpha\}$ can be obtained by fitting $\rho_{\dot{x}}^{(\varepsilon)}(\tau)$ at long lag times, $\varepsilon$. However, determining how to extract and understand the relaxation parameters requires a microscopic stochastic model for SM transport, which we discuss in the next section.

**Microscopic Dynamics and Numerical Modeling**

**Generalized Langevin equation and fractional Gaussian noise:** The dynamics of a single particle subjected to a potential $V(x,t)$ with velocity $\dot{x}(t)$ over a time window [0,$t$) is governed by the generalized Langevin equation (GLE) (29),

$$m\ddot{x}(t) = -\frac{\partial V(x,t)}{\partial x} - \int_0^t \eta(t-t')\dot{x}(t')dt' + \xi(t) \qquad (3)$$

Here, $\xi(t)$ is the random thermal force exerted on the particle by the environment at temperature $T$. Its autocorrelation $\langle\xi(t)\xi(t')\rangle$ is related to the frictional kernel $\eta(t)$ via the fluctuation-dissipation theorem (FDT) (30)

$$\langle\xi(t)\xi(t')\rangle = k_B T\eta|t-t'|. \qquad (4)$$

$\xi(t)$ arises from the initial positions and momenta, which are sampled from a canonical distribution, resulting in stationary Gaussian statistics (31, 32). The power spectral density, $S(\omega)$, of the thermal noise - defined as the Fourier transform of the autocorrelation of $\xi(t)$ - is related to the spectral bath density, $J(\omega)$, which characterizes system–bath coupling and encodes the bath's influence on particle dynamics. Specifically,

$$S(\omega) = \int_{-\infty}^{\infty} \langle \xi(t)\xi(0)\rangle e^{-i\omega t} dt = \frac{2k_B T J(\omega)}{\omega}. \tag{5}$$

A physically meaningful form of the spectral density $J(\omega) = \eta_\alpha |\sin(\pi\alpha/2)|\omega^\alpha$ corresponds to a power-law friction kernel (22, 32–35):

$$\eta(t) = \frac{\eta_\alpha}{\Gamma(1-\alpha)} \frac{1}{t^\alpha} \tag{6}$$

and gives rise to $1/f^{1-\alpha}$ thermal noise, also known as fractional Gaussian noise (fGN) (14, 16, 22, 33–38). Here $\Gamma()$ is the Gamma function. Such friction kernels have previously been used to describe the physical behavior of disordered media (4). In this study, we approximated our polymeric system as a heat bath that drives the molecular probe through random fGN thermal kicks (Fig. S2a). In the limit $\alpha \to 1$, the kernel reduces to $\delta(t)$, corresponding to a memoryless Ohmic bath that responds instantaneously, leading to Stokes friction. In contrast, for a sub-Ohmic bath ($0 < \alpha < 1$) the power-law friction kernel captures the viscoelastic response of complex media with long-range temporal memory, in which elastic components of the viscoelastic force tend to push the particle back towards its prior position (4).

In high-viscosity media, friction dominates over inertia, and the overdamped limit $m \to 0$ becomes relevant. Then, the GLE dynamics with a restoring potential $V(x,t)$ relaxes towards equilibrium. Under stationary initial conditions, the coordinate autocorrelation function becomes: $\langle x(t).x(t')\rangle = \langle x_T^2\rangle \theta(|t-t'|)$, where $\langle x_T^2\rangle$ is the thermal equilibrium variance and $\theta(t)$ is the relaxation function [SI Appendix]. For a particle in a parabolic potential with a spring constant $\kappa$ under $1/f^{1-\alpha}$ thermal noise, relaxation follows the Mittag-Leffler function, $\theta(t) = E_\alpha\left[-\left(\frac{t}{\tau_D}\right)^\alpha\right]$, where $\tau_D = (\eta_\alpha/\kappa)^{1/\alpha}$ is a characteristic translational relaxation timescale of the system [SI Appendix] (23, 33). In the limit $\alpha \to 1$, relaxation reduces to the familiar single exponential decay, whereas for $\alpha < 1$ the relaxations are broadly distributed. Near $\tau_D$, the Mittag-Leffler function is well approximated by a stretched exponential also known as the Kohlrausch–Williams–Watts (KWW) law, $\theta(t) = e^{[-(t/\tau_D)^\alpha]}/\Gamma(1+\alpha)$, which will be used to represent $\theta(t)$ unless stated otherwise (4, 39, 40).

**Anomalous transport:** In a sub-Ohmic bath, the fGN thermal kicks act to restore the particle toward its earlier positions such that past and current motions are anti-correlated. This memory effect slows down the transport relative to Markovian systems with a purely viscous response. Quantitatively, for a sub-Ohmic bath with fGN, the TAMSD for a particle in a parabolic trap is given by [SI Appendix]:

$$\langle \Delta x^2(\varepsilon) \rangle = \frac{2k_B T}{\kappa}(1 - \theta|\varepsilon|) \tag{7}$$

In the free-particle limit ($\kappa \to 0$), this reduces to the well-known sub-diffusive expression (16, 33, 39)

$$\langle \Delta x^2(\varepsilon) \rangle = \frac{2D_\alpha(\varepsilon)^\alpha}{\Gamma(1+\alpha)}. \tag{8}$$

$D_\alpha$ is related to the fractional friction coefficient $\eta_\alpha$ via $D_\alpha = k_B T/\eta_\alpha$. In both the trapped and free cases, the presence of long memory implies that the particle takes longer to explore a given space than in the diffusive case ($\alpha \to 1$). It is worth noting that for a trapped particle, the TAMSD reaches a plateau at long times ($t \gg \tau_D$), whereas for a free particle, the TAMSD continues to grow sub-linearly with time (Fig. S2b).

**Anti-correlated displacement rate correlation:** For a sub-Ohmic bath, the thermal kicks are temporally correlated and the resulting increments are not independent: a positive increment is likely to be followed by a negative one, in contrast to a free particle in an Ohmic bath (26). This anti-persistent stochastic mechanism leads to a negative displacement correlation and reflects the memory effects intrinsic to the sub-Ohmic bath. $\rho_{\dot{x}}^{(\varepsilon)}(\tau)$ for a particle in a parabolic trap subject to a viscoelastic kernel becomes [SI Appendix]:

$$\rho_{\dot{x}}^{(\varepsilon)}(\tau) = \frac{(2\theta|\tau| - \theta|\tau + \varepsilon| - \theta|\tau - \varepsilon|)}{2(1 - \theta|\varepsilon|)}. \tag{9}$$

In the free particle limit ($\kappa \to 0$), this reduces to well-known results reported by Jeon and Metzler (16):

$$\rho_{\dot{x}}^{(\varepsilon)}(\tau) = \frac{(|\tau + \varepsilon|^\alpha + |\tau - \varepsilon|^\alpha - 2\tau^\alpha)}{2\varepsilon^\alpha}. \tag{10}$$

$\rho_{\dot{x}}^{(\varepsilon)}(\tau)$ typically reaches its minimum at $\tau = \varepsilon$. Importantly, $\Delta x(\varepsilon)/\varepsilon$ at a given lag $\varepsilon$ constitutes an $\rho_{\dot{x}}^{(\varepsilon)}(\tau)$ function, which provides a more precise means of extracting dynamical characteristics relative to TAMSD, especially since early TAMSD values can be prone to effects from localization errors. Interestingly, the presence of a potential trap gradually increases the magnitude of the negative correlation with increasing $\varepsilon$, which helps to distinguish between a trapped and free particle system (27).

**Relaxation dynamics from displacement-fluctuation:** Fig. 4a shows $\overline{\rho_{\dot{x}}^{(\varepsilon)}(\tau)}$ curves, obtained from averaging multiple $\rho_{\dot{x}}^{(\varepsilon)}(\tau)$ and fit to the stochastic microscopic model (Eq. 9) at $\varepsilon > \tau_c$. The fits show excellent agreement with the experimental data at all three temperatures, suggesting that the transport behavior near $T_g$ is well captured by the stochastic microscopic framework presented. The microscopic model yields, $\tau_D = 0.11$s, $4.5$s, and $62.3$s and $\alpha = 0.45, 0.35$ and $0.38$ for 380.6K, 377.6K and 374.5K, respectively. Physically, $\tau_D$ represents the characteristic time over which the particle explores the parabolic potential under fGN, balancing the viscoelastic friction ($\eta_\alpha$) and the restoring force ($\kappa$). It is worth noting that the thermal relaxation timescales of the system derived from probe translation are in accord with those obtained from its rotational dynamics (9).

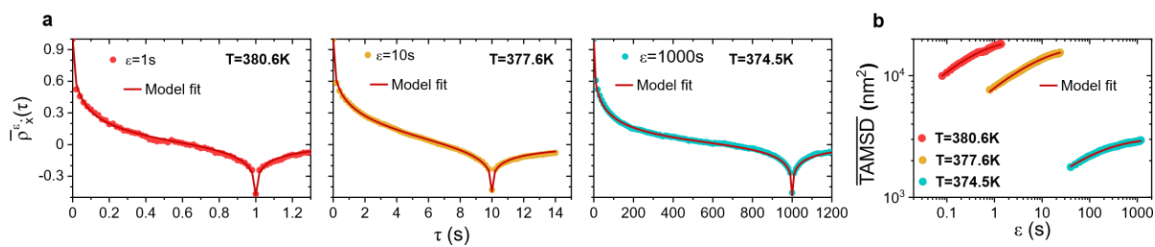

**Fig. 4.** Comparison of experimental displacement rate correlation functions and TAMSDs with the stochastic model discussed in the text. Shown are the experimental averages of (a) $\rho_{\dot{x}}^{(\varepsilon)}(\tau)$ and (b) TAMSD curves of many individual probes (dots) and the fits to these $\overline{\rho_{\dot{x}}^{(\varepsilon)}(\tau)}$ and $\overline{\text{TAMSD}}$ data (lines). In (a), left, middle, and right panels show data at 380.6K, 377.6K, and 374.5K, respectively. In (b), the first five data points of the $\overline{\text{TAMSD}}$ curves were excluded from fitting to minimize effects of possible artifacts due to localization inaccuracy.

Fig. 4b shows $\overline{\langle \Delta x^2(\varepsilon) \rangle}$ curves computed by averaging TAMSDs of individual probes shown in Fig. 3a. When $\alpha$ values obtained from $\overline{\rho_{\dot{x}}^{(\varepsilon)}(\tau)}$ are used to fit $\overline{\langle \Delta x^2(\varepsilon) \rangle}$ via Eq. 7, the model again results in a good fit to the experimental data, suggesting consistency between the transport behavior and the presented model. The downward curvature indicates the onset of probe confinement. Fig. S3 (SI Appendix) shows the relaxation functions, $\overline{\theta|\tau|}$, obtained from the autocorrelation of experimental $x(t)$ fluctuations from multiple probes. At elevated temperature (376.6K and 380.6K), the relaxations decay below 0.15, indicating that experimental timescale is sufficient to capture the equilibrium fluctuations. However, at the lower temperature (374.5K), the relaxation remains above 0.20, indicating long decay times for a subset of probes compared to the trajectory length. This observation is consistent with the non-vanishing third-order moment shown in Fig. 1c, which reflects the long-lived dynamic heterogeneity upon approaching $T_g$. Due to this and short-time localization uncertainty, we refrain from retrieving the relaxation parameters $\{\tau_D, \alpha\}$ from the SM $\theta|\tau|$.

**Numerical modeling of heterogeneous transport:** A key feature of the GLE dynamics (Eq. 3) is that the dissipative force $F_{diss} = \int_0^t \eta(t-t')\dot{x}(t')dt'$ remains linear despite its temporally non-local dependence on past velocity. In a parabolic potential, the GLE yields Gaussian statistics for the position fluctuations (39). We numerically simulate stochastic behavior to obtain trajectories exhibiting sub-diffusive transport with anti-correlated displacements and non-Gaussian $P(\Delta x, \varepsilon)$ (see Methods). In these numerical simulations, we assume that the sub-diffusive parameters $\{D_\alpha, \alpha\}$ remain constant, corresponding to Gaussian statistics, as are seen to exist over short timescales (Fig. 2c). To capture the heterogeneous environment, each realization mimics such a time-segment and assigns a random pair $\{D_\alpha, \alpha\}$ sampled from a log-normal and normal distribution, respectively. In our current implementation, the fGN characteristics and the confinement strength, $\kappa$, are defined at the ensemble level for each temperature, while the probe-to-probe variation is introduced by the sub-diffusive parameters $\{D_\alpha, \alpha\}$ to match the experimental observations. In principle, variable time-segments can be chosen for different SM trajectories; however, since our simulation does not explicitly model temporal exchange, this is unnecessary. We simulate 5000 time-points with resolution ($\Delta t$) adjusted to the experimental frame rate and broaden the spread of $\{D_\alpha, \alpha\}$ to account for temporal variation.

Fig. 5 shows numerical results, which reproduce the essential dynamical behavior of the experimental results near $T_g$. Non-Gaussian behavior arises only when transport parameters are distributed. To account for the heterogeneity, we broaden the distribution of $\{D_\alpha, \alpha\}$; the broader this distribution, the stronger the non-Gaussianity. In the simulation, we set $\langle \alpha \rangle = 0.4$ at 374.5K and 377.6K and 0.45 at 380.6K - values close to those observed experimentally. The median of $D_\alpha$ was adjusted to align with the peaks of the experimental displacement distributions, yielding $logD_\alpha = 2.38$ nm$^2$s$^{-\alpha}$, $3.25$ nm$^2$s$^{-\alpha}$, and $4.1$ nm$^2$s$^{-\alpha}$ respectively, which are only slightly shifted from the experimental values. To account for temporal heterogeneity, $\sigma_\alpha$ was fixed at 0.2, approximately twice the experimental value, while the FWHM of $logD_\alpha$ was tuned to reproduce the experimental non-Gaussian behavior. The numerical simulation reproduces the experimental displacement distributions well across all three temperatures (Fig. 5a). The FWHM of the $logD_\alpha$ distribution is required to be 2.0, 2.6, and 2.7 for 380.6K, 377.6K and 374.5K, respectively. The good agreement between the experimental and simulated results indicates that the minimal stochastic numerical model can faithfully capture the displacement distribution. The broad distribution of $\{D_\alpha, \alpha\}$ required in the numerical simulation to account for the experimental displacement distribution implies the presence of diverse heterogeneous environments which increase upon approaching $T_\mathrm{g}$.

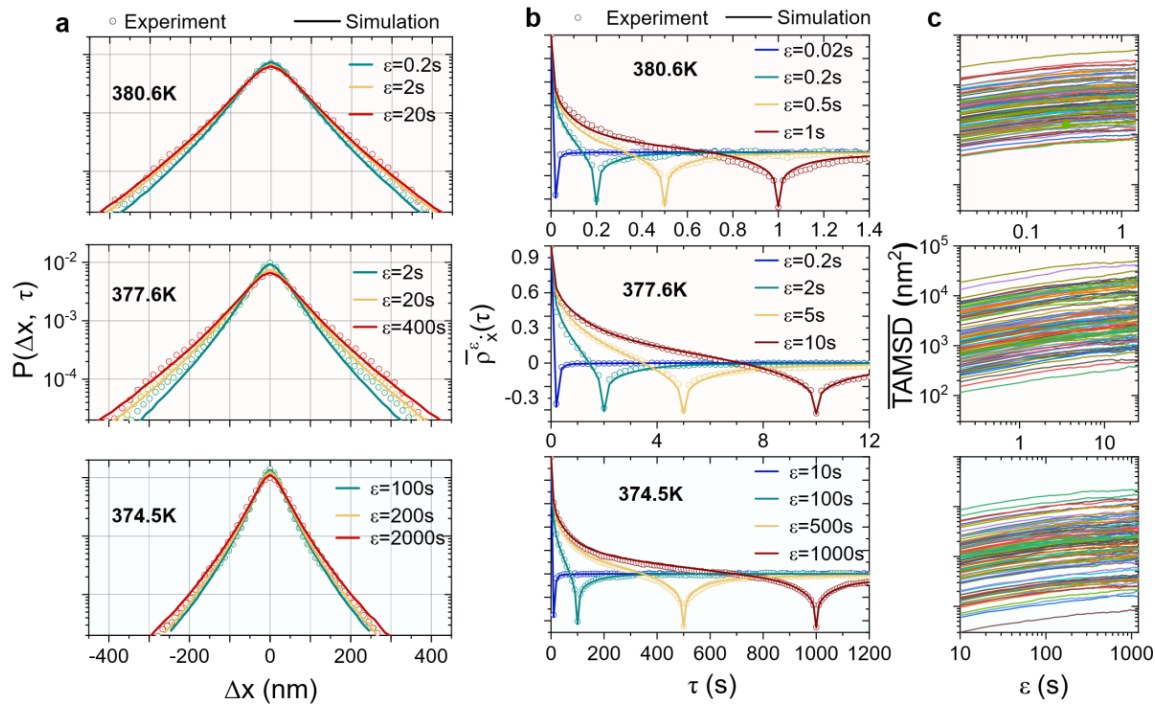

**Fig. 5.** Numerical simulation of the GLE with a power-law friction kernel in a harmonic potential with heterogeneous transport. (a) Probability density functions, $P(\Delta x, \varepsilon)$ of the displacement distributions of SM probes. The experimental data are shown as circles, while the simulated data are shown as solid lines. Results are shown for three $\varepsilon$ values. (b) $\overline{\rho_{\dot{x}}^{(\varepsilon)}(\tau)}$ as obtained from multiple trajectories. The experimental data are shown as circles, and the simulated data, which represent the average of 1000 realizations, are shown as solid lines. Results are shown for four $\varepsilon$ values at each temperature, as indicated by different colors. (c) One hundred representative TAMSDs from the simulation, demonstrating the broad spread of sub-diffusive dynamics necessary to recapitulate the experimental dynamic heterogeneity.

Fig. 5b shows that the numerical simulation produces anti-correlated displacements, evident from the negative peak of $\overline{\rho_{\dot{x}}^{(\varepsilon)}(\tau)}$, as in the experimental results. The negative peak is consistent with the thermal sub-Ohmic bath with anti-persistent random thermal kicks. As presented in Fig. 5b, the chosen sets of transport parameters capture the negative displacement correlation with remarkable agreement with experimental observation. The numerical $\tau_D$ extracted from fitting $\overline{\rho_{\dot{x}}^{(\varepsilon)}(\tau)}$ is in close accord with the experimental $\tau_D$ (fit not shown). This reflects that the behavior from hundreds of heterogeneous trajectories reproduces the ensemble behavior. Importantly, the numerical simulation also exhibits a progressive increase in the absolute value of the negative correlation peak with increasing $\varepsilon$ due to the confining potential, which is consistent with the experimental observations. The probe-to-probe variation of the translational transport dynamics is reflected in the TAMSDs, as shown in Fig. 5c for 100 representative trajectories. The spread of the TAMSDs appears broader than that observed experimentally (Fig. 3a), which was chosen intentionally to match the experimental displacement distribution from all available molecules. This broader distribution arises ostensibly from the fact that we model the temporal exchange by increasing the breadth of $\{D_\alpha, \alpha\}$.

**Discussion**

Polymer segmental dynamics near $T_\mathrm{g}$ give rise to diverse probe transport behaviors. At short timescales, sub-diffusive translational dynamics are evident, apparently reflecting the viscoelastic nature of the rubbery polymer. At longer timescales, the probes are apparently locally confined, presumably by the surrounding polymer chains. Our results show that the probes exhibit non-Gaussian displacement distributions; however, over short timescales, their position fluctuations remain Gaussian. Local equilibrium-like position fluctuations based on the GLE with a sub-Ohmic thermal bath, along with a numerical model accounting for spatiotemporal heterogeneity, successfully captures the experimental displacement distributions and other essential transport behaviors. In the model, probes explore spatiotemporally heterogeneous environments, thereby experiencing multiple independent sub-diffusive processes, which underpins the spread of mean-squared displacements across individual probes near $T_g$ (13). We note that the non-Gaussian behavior can be captured using naive GLE dynamics with randomly switching microenvironments that represent exchange dynamics, but such an approach may lack thermodynamic consistency. On the other hand, theoretical frameworks describe the dynamical behavior near and above $T_g$ within an equilibrium framework, as only below $T_g$ do amorphous systems fall out of equilibrium (3, 41, 42). Under this hypothesis, intermittent transitions between dynamically cooperative domains must satisfy detailed balance to ensure a stationary quasi-equilibrium (6). In a different approach, the diffusivity itself is modeled as a continuous, stochastic, time-varying process (3). While this *diffusing diffusivity* model captures many features of non-Gaussian fluctuations, a fully consistent microscopic GLE derivation that preserves detailed balance remains an open challenge (3, 13, 43–45).

We highlight that the confinement length scale and translational relaxation timescales of the probes measured experimentally near $T_g$ are unexpected. Conventionally, the non-Gaussian fat tail of the probe displacement distribution is attributed to microscopic structural relaxation, known as $\alpha$-relaxation, which characteristically drives probe cage escape (1, 19–21). In such cases, the MSD typically shows an initial plateau, reflecting its cage dynamics associated with $\beta$-relaxation, followed by a gradual increase in slope, approaching diffusive behavior due to $\alpha$-relaxation (1, 20, 28). In contrast, we observe a decreasing MSD slope, suggestive of a trap mechanism, where non-Gaussian tails are prevalent at timescales longer than the $\beta$-relaxation time of the system. This confining potential, with a length scale 15-35 nm in polystyrene near $T_g$ - as inferred from the plateau of the $\overline{\langle \Delta x^2(\varepsilon) \rangle}$ curves - is unexpected and may originate from polymer entanglements and/or probe-polymer intermolecular interactions. SM tracking experiments in unentangled polymers or small-molecule glass formers may help clarify the origin of this confinement. Meanwhile, the agreement between the translational relaxation and rotational correlation times suggests a common origin in the thermal segmental motions of the polymer chains. Together, these findings underscore the need for a deeper understanding of relaxation and transport in polymer glass formers near $T_g$.

In conclusion, we performed a detailed analysis of SM trajectories to understand dynamics near $T_g$. We provide a microscopic framework based on the GLE to explain the experimentally observed probe translational motion. The polymer is modeled as a sub-Ohmic bath generating correlated thermal noise that causes anomalous sub-diffusive transport, while surrounding polymer segments generate the confining potential, which together give rise to the characteristic probe relaxation dynamics. To incorporate heterogeneity, the transport parameters are sampled from well-defined distributions that reflect the probe's exploration of many independent sub-diffusive processes over time. Although SM translational tracking is widely used to study structurally heterogeneous soft matter and biological systems, local equilibrium picture-based analysis of transport and relaxation mechanisms remains to be explored to examine dynamic heterogeneity in molecular systems near $T_g$. We emphasize that the accumulation of large experimental data sets across a variety of molecular and polymer glass formers holds significant potential to disentangle the causes and consequences of dynamic heterogeneity. Our microscopic framework offers a new perspective for investigating experimentally observed translational dynamics in a variety of complex systems and is particularly well suited to elucidating glassy dynamics near $T_g$.

**Materials and Methods**

**Sample preparation:** Here we analyze probe dynamics in polystyrene at three temperatures: 380.6K, 377.6K and 374.5K. The data sets at 380.6K and 377.6K were previously reported (11) and re-analyzed here while the 374.5K dataset was newly recorded. The detailed sample preparation can be found in Ref. (11), and the new measurements followed the same sample preparation protocol, with the only difference being the data collection procedure. In brief, atactic polystyrene ($M_w$ = 168 kg/mol, PDI = 1.05; Polymer Source) was reprecipitated in hexane or toluene and then dissolved in toluene at 3.5–4.0 wt%. This solution was photobleached under a high-power LED setup for 3-6 days to minimize fluorescent impurities. The fluorescent probe, N,N′-dipentyl-3,4,9,10-perylenedicarboximide (pPDI; Sigma-Aldrich) was dissolved in toluene and added to an aliquot of PS solution resulting in final probe concentration of $\approx 10^{-11}$M. Approximately 30 μL of this solution was spin coated at 3000 rpm onto a 7mm × 7mm silicon wafer (pre-treated with piranha solution, 2:1 $H_2SO_4$:$H_2O_2$). The resulting films were at least 200nm thick as measured by ellipsometry, ensuring that observed dynamics reflect bulk behaviors. Films were annealed in a vacuum chamber (Janis ST-500) at 390K and 4.5mTorr $\geq 12h$, then cooled to the target temperature and equilibrated for several hours prior to SM imaging.

**Data Collection:** All SM data was collected on home-built wide-field fluorescence microscope setups. The procedure for the previously collected data can be found in Ref. (11) and the new measurements followed similar protocols. Briefly, a 532 nm laser (CL532-200-L0; CrystaLaser LC) was coupled to a multimode fiber (Newport, F-MCB-T-1FC) that was shaken by a speaker to ensure a homogeneous illumination field. The laser was focused at the back focal plane of the objective lens (Olympus, MPLAPON, 100X, NA = 0.95, WD = 0.3 mm) which was housed inside a vacuum- and a temperature-controlled chamber, with laser excitation power ≈14mW measured just before the objective. Fluorescence emitted from the sample was collected in the epi-direction by the same objective and passed through an appropriate dichroic mirror (Semrock, LPD02-532RU) followed by bandpass and long pass filters before imaging on an electron-multiplying CCD camera (EMCCD; Andor iXon DV887). A Wollaston prism was placed before the EMCCD which split the light into two orthogonal polarization channels, ensuring measurement consistency with the previous report (11).

The earlier measurements at 380.6K and 377.6K were performed at frame rates of 50Hz and 5Hz, respectively, with exposure times equal to their corresponding frame intervals ($\Delta t$). In the earlier work [Ref.(11)], data was also collected at lower temperatures. However, we realized that lower frame rates were required to obtain good fits to Eq. 7 and Eq, 9, such as those seen in Fig. 4. Thus, new measurements were conducted at 0.1 Hz with a set exposure time of 2s at 374.5K. Additionally, trajectory length was set to $\geq 5000$ frames ($200\tau_{fit}$, see Data Analysis) to obtain reliable statistical behavior (7, 8).

**Data Analysis:** The SM movies were analyzed to generate position trajectories, $\{x_i(t), y_i(t)\}$ using the procedure outlined in Ref. (10). For simplicity, we denote either coordinate as $x(t)$, representing motion along a single dimension. Briefly, movies collected in two orthogonal polarization channels were drift-corrected and combined into a single channel to track the SMs, referred to as features. These features were localized using the ImageJ Particle Tracker that employed intensity-weighted centroids to obtain individual SM coordinates (10, 46). Because SMs may undergo photoblinking or orientation towards the optical axis, features can be lost, and trajectories can be truncated. To avoid this, shorter trajectories corresponding to the same molecule were stitched together using a hierarchical clustering method, producing longer trajectories in which missing frames were marked as NaN. This is the main difference in trajectory generation between the earlier report (11) and the more recent procedure (10), which we adopt in the present work. Trajectories with more than 70% missing frames were discarded prior to performing any second and higher order correlation analyses to ensure statistically meaningful results.

For translational dynamics, several complementary approaches were employed, and many analysis details are provided in the experimental results section. Fits to different translational observables (Eq.7 and Eq.9) typically involve a stretched exponential, $C(0)e^{[-(t/\tau_D)^\alpha]}$, where $C(0)$ is a prefactor, $\tau_D$ is relaxation time and $\alpha$ is the stretching (or anomalous diffusion) exponent. Parameter bounds were chosen similar to those in our previous work: $0.1 < C(0) < 2.0$; $0.03 < \alpha < 1.0$ and $\Delta t < \tau_D <$ movie length(s). Short-time lag points of TAMSDs were fit with Eq. 8 to obtain the sub-diffusive parameters $\{D_\alpha, \alpha\}$. To ensure robust statistics, all trajectories with initial positive slopes on the log-log scale of TAMSD were retained. All fits were performed using MATLAB's nonlinear least-squares routine with $R^2 > 0.9$ for SMs and 0.99 for the quasi-ensemble. The change-point analysis of trajectory fluctuations associated with Fig. 2c, based on moving variance, was performed using MATLAB's built-in algorithm. For rotational dynamics, the fluorescence intensity of each SM in each polarization channel, $\{I_s(t), I_p(t)\}$, were simultaneously monitored to obtain the characteristic rotational correlation time scales and stretching exponent from linear dichroism, $LD(t) = [I_s(t) - I_p(t)]/[I_s(t) + I_p(t)]$. The autocorrelation of $LD(t)$ was fit to $C(0)e^{[-(t/\tau_{fit})^\beta]}$ where the average rotational correlation time of SM is given by $\tau_c = (\tau_{fit}/\beta)\Gamma(1/\beta)$. The details of this analysis can be found in our earlier work (9–11). All image analyses were carried out in ImageJ, trajectory clustering in Python and statistical data analysis in MATLAB 2022a.

# Appendix

## I. Power-law governed Non-Markovian Microscopic Dynamics

**GLE solution and single particle trajectory:** A comprehensive treatment of the Generalized Langevin equation (GLE) and its two-time correlation functions is available in Ref (39). Here, we derive the essential aspects relevant to our model and subsequent analysis. In the overdamped limit, the GLE for a particle in a parabolic potential reads:

$$\int_0^t \eta(t-t')\dot{x}(t')dt' = -\kappa x(t) + \xi(t). \tag{A1}$$

The particle trajectory $x(t)$ driven by thermal noise $\xi(t)$ can be obtained by Laplace transforming both sides. In the Laplace domain, the GLE reads:

$$\tilde{\eta}(s).[s\tilde{x}(s) - x(0)] = -\kappa \tilde{x}(s) + \tilde{\xi}(s). \tag{A2}$$

Solving, $\tilde{x}(s)$, we get:

$$\tilde{x}(s) = \frac{\tilde{\eta}(s)}{s\tilde{\eta}(s) + \kappa} x(0) + \frac{1}{s\tilde{\eta}(s) + \kappa} \tilde{\xi}(s) \tag{A3}$$

Here, $\tilde{x}(s)$, $\tilde{\eta}(s)$ and $\tilde{\xi}(s)$ are the Laplace transform of $x(t)$, $\eta(t)$ and $\xi(t)$, respectively. Taking the inverse Laplace transform gives the real-time trajectory:

$$x(t) = x(0)\theta(t) + \int_0^t \chi(t-t')\xi(t') \, dt' \tag{A4}$$

This shows that the particle's response to thermal fluctuations is temporally nonlocal - past noise $\xi(t')$ contributes to the present position $x(t)$. The first term describes the particle's deterministic relaxation from an arbitrary initial position, $x(0)$, towards equilibrium modulated by $\theta(t)$. With the Laplace transform, $\tilde{\theta}(s) = \frac{\tilde{\eta}(s)}{s\tilde{\eta}(s) + \kappa}$. The second term captures the response to stochastic thermal fluctuations, with $\tilde{\chi}(s) = \frac{1}{s\tilde{\eta}(s) + \kappa}$ the Laplace transform of the coordinate response function. The response function, $\chi(t)$, also describes linear response to an external impulse - for instance the dielectric susceptibility in glassy media, and $\theta(t)$ becomes the relaxation decay.

For the power-law memory kernel, $\eta(t) = \eta_\alpha t^{-\alpha}/\Gamma(1-\alpha)$, its Laplace transform becomes $\tilde{\eta}(s) = \eta_\alpha s^{\alpha-1}$. Substituting this into the expressions for $\theta(t)$ and $\chi(t)$ yields:

$$\theta(t) = E_\alpha\left[-\left(\frac{t}{\tau_D}\right)^\alpha\right] \text{ and } \chi(t) = \frac{t^{\alpha-1}}{\eta_\alpha} E_{\alpha,\alpha}\left[-\left(\frac{t}{\tau_D}\right)^\alpha\right]. \tag{A5}$$

where $E_\alpha(\cdot)$ and $E_{\alpha,\beta}(\cdot)$ are the one- and two-parameter Mittag–Leffler functions, respectively and $\tau_D = (\eta_\alpha/\kappa)^{1/\alpha}$ is the relaxation timescale (39). The Mittag–Leffler function, $E_{\alpha,\beta}(-z) = \sum_{n=0}^{\infty} \frac{(-1)^n z^n}{\Gamma(n\alpha+\beta)}$ generalizes the exponential function and appears naturally in systems with power-law memory. For $\beta = 1$, $E_{\alpha,\beta}(\cdot)$ becomes $E_\alpha(\cdot)$. In the frequency domain, $\chi(t)$ for this viscoelastic kernel becomes the Cole-Cole spectrum $\chi(\omega)$ (4).

In the free-particle limit $\kappa \to 0$, the trajectory reduces to

$$x(t) = x(0) + \frac{1}{\eta_\alpha \cdot \Gamma(\alpha)} \int_0^t (t-t')^{\alpha-1} \xi(t')\, dt' \tag{A6}$$

This convolution integral is formally recognized as the Riemann–Liouville fractional integral of order $\alpha$ denoted as $I^\alpha \xi(t)$ (34, 36) such that

$$x(t) = x(0) + \frac{1}{\eta_\alpha \cdot} I^\alpha \xi(t) \text{ where } I^\alpha \xi(t) = \frac{1}{\Gamma(\alpha)} \int_0^t (t-t')^{\alpha-1} \xi(t')\, dt'. \tag{A7}$$

This is analogous to the integral representation of fractional Brownian motion (fBM), which is also constructed via fractional integration. The resulting stochastic equation is often referred to as the fractional Langevin equation (FLE), highlighting the nonlocal, memory-dependent dynamics where the memory strength is governed by the exponent $\alpha$ (14, 16, 36).

**Two-time Position Correlation Function:** Since $x(t)$ is itself a convolution over $\xi(t)$, its autocorrelation requires the double Laplace transform of the noise autocorrelation. Exploiting the time symmetry of the memory kernel $\eta|t - t'|$, the integral splits into two domains yielding (48)

$$\langle \tilde{\xi}(s) \cdot \tilde{\xi}(s') \rangle = k_B T \int_0^\infty \int_0^\infty e^{-st} e^{-s't} \eta|t-t'|\, dt\, dt' = k_B T \left[ \frac{\tilde{\eta}(s) + \tilde{\eta}(s')}{s+s'} \right] \tag{A8}$$

This relation helps to construct the position autocorrelation $\langle \tilde{x}(s)\tilde{x}(s') \rangle$ in the Laplace domain. Assuming initial zero mean and variance $\langle x(0)^2 \rangle$, the position autocorrelation in the time domain becomes

$$\langle x(t) \cdot x(t') \rangle = \langle x_T^2 \rangle \theta(|t-t'|) + [\langle x(0)^2 \rangle - \langle x_T^2 \rangle]\theta(t)\theta(t') \tag{A9}$$

Here, $\langle x_T^2 \rangle = \frac{k_B T}{\kappa}$ is the thermal equilibrium variance. This result follows from a known decomposition in the literature, where the stochastic part contributes a term proportional to $\langle x_T^2 \rangle [\theta(|t-t'|) - \theta(t)\theta(t')]$. Under equilibrium initial conditions, $\langle x(0)^2 \rangle = \langle x_T^2 \rangle$, the result simplifies to: $\langle x(t) \cdot x(t') \rangle = \langle x_T^2 \rangle \theta(|t-t'|)$. This is the Onsager regression: the relaxation function becomes the normalized stationary autocorrelation function.

In the free particle limit ($\kappa \to 0$) and assuming $x(0) = 0$, the stochastic fluctuation leads to the following two point correlation(35):

$$\langle x(t).x(t') \rangle = \frac{k_B T}{\eta_\alpha \Gamma(1+\alpha)} [t^\alpha + t'^\alpha - |t'-t|^\alpha] \tag{A10}$$

This second-order structure highlights the connection between FLE dynamics and that of fBM with long-range temporal correlations. The two-point correlations are subsequently utilized for the calculation of time-average MSD and the other correlation functions. Note that if the increments of fBM are used as the noise source, the resulting noise covariance scales as $t^{2-2H}$ where $H$ is the Hurst exponent (14, 16).

**Correlation functions and TAMSD:** To connect theory with single-molecule trajectory data, we analyze the statistical properties of lagged displacements. The lagged displacement over m-lag interval $\varepsilon = m\Delta t$ is defined as, $\Delta x(\varepsilon) = x(t+\varepsilon) - x(t)$. The autocorrelation function of the time series is

$$C_{\Delta x}^{(\varepsilon)}(\tau = n\Delta t) = \langle \{x(t+\tau+\varepsilon) - x(t+\tau)\}\{x(t+\varepsilon) - x(t)\}\rangle. \tag{A11}$$

Using the position correlation function $\langle x(t).x(t')\rangle$ at thermal equilibrium, this evaluates to

$$C_{\Delta x}^{(\varepsilon)}(\tau) = \frac{k_B T}{k} (2\theta|\tau| - \theta|\tau+\varepsilon| - \theta|\tau-\varepsilon|). \tag{A12}$$

For a free particle ($\kappa \to 0$), this reduces to

$$\lim_{k\to 0} C_{\Delta x}^{(\varepsilon)}(\tau) = \frac{k_B T}{\eta_\alpha \Gamma(1+\alpha)} (|\tau+\varepsilon|^\alpha + |\tau-\varepsilon|^\alpha - 2|\tau|^\alpha). \tag{A13}$$

In particular at $\tau = 0$; $C_{\Delta x}^{(\varepsilon)}(\tau)$ yields the time-averaged mean-squared displacement (TAMSD):

$$C_{\Delta x}^{(\varepsilon=m\Delta t)}(0) = \langle [x(t+\varepsilon)-x(t)]^2 \rangle = \langle [\Delta x(\varepsilon)]^2 \rangle = \frac{2k_B T}{\kappa}(1-\theta|\varepsilon|). \tag{A14}$$

In the free particle limit ($\kappa \to 0$), one obtains the well-known pure sub-diffusive TAMSD

$$\lim_{k\to 0}\langle [\Delta x(\varepsilon)]^2 \rangle = \frac{2k_B T}{\eta_\alpha \Gamma(1+\alpha)}(\varepsilon)^\alpha. \tag{A15}$$

Defining the displacement per unit time as $\Delta x(\varepsilon)/\varepsilon$, the normalized autocorrelation function of the finite-time displacement rate (velocity) becomes:

$$\rho_{\dot{x}}^{(\varepsilon)}(\tau) = \frac{C_{\dot{x}}^{(\varepsilon)}(\tau)}{C_{\dot{x}}^{(\varepsilon)}(0)} = \frac{(2\theta|\tau| - \theta|\tau+\varepsilon| - \theta|\tau-\varepsilon|)}{2(1-\theta|\varepsilon|)}. \tag{A16}$$

In the free particle limit ($\kappa \to 0$), this reduces to the similar result reported by Jeon and Metzler:

$$\lim_{k \to 0} \rho_{\dot{x}}^{(\varepsilon)}(\tau) = \frac{(|\tau + \varepsilon|^{\alpha} + |\tau - \varepsilon|^{\alpha} - 2\tau^{\alpha})}{2\varepsilon^{\alpha}} \quad (A17)$$

Recently the quantity $\rho_{\dot{x}}^{(\varepsilon)}(\tau)$ has been used to characterize various stochastic processes (15, 16, 18, 24). Evaluating at discrete step $\tau = n\Delta t$ for a particular $\varepsilon = m\Delta t$, Eq. A17 becomes

$$\lim_{k \to 0} \rho_{\dot{x}}^{(m)}(n) = \frac{(|n + m|^{\alpha} + |n - m|^{\alpha} - 2n^{\alpha})}{2m^{\alpha}} \quad (A18)$$

This normalized correlation is scale-invariant and useful for characterizing memory and is similar to that obtained in Ref. (16) with $m = 1$. For the sub-Ohmic bath $(0 < \alpha < 1)$, the correlation reaches a minimum (most negative) when $n = m$, given by $\rho_{\dot{x}}^{(k=n)}(n) = (2^{\alpha} - 2)/2$ (16, 18). For free particles, when $n \gg m$ the correlation decays to zero, reflecting uncorrelated displacements at long times. The anti-correlation of displacements before the limiting behavior at large $n$ is a hallmark of anti-persistent motion of $1/f^{1-\alpha}$ thermal noise (15, 18, 24, 26).

## II. Numerical Solution of the GLE:

To generate single particle trajectories, we numerically solve the microscopic dynamics described by the GLE with a power-law frictional kernel in Eq. A4. It is important to note that $\xi(t)$ is colored Gaussian noise, not white noise. Furthermore, the random trajectory generation involves a convolution of $\xi(t)$ with the response function, $\chi(t)$. For numerical stability, we precompute the integral of $\chi(t)$ to obtain the discrete cumulative response function, which is then used to generate the discrete convolution weights for $\chi(t)$.

**Fractional Gaussian noise generation:** We construct the correlated thermal noise following the procedure outlined in Refs. (18, 26, 39). Briefly, we approximate the $1/f^{1-\alpha}$ thermal noise as the sum of many independent Ornstein-Uhlenbeck (OU) processes, each governed by a stiffness $k_i$ and relaxation rate $\nu_i$, and defined by the stochastic differential equation:

$$\dot{\zeta}_i(t) = -\nu_i \zeta_i(t) + \sqrt{2k_B T k_i \nu_i} \rho_i(t) \quad (B1)$$

where $\rho_i$ are the independent Gaussian white noise satisfying $\langle \rho_i(t)\rho_j(t') \rangle = \delta_{ij}\delta(t - t')$. Note that the initial $\zeta_i(0)$ are sampled from an unbiased zero mean Gaussian random distribution with variance $\langle \zeta_i^2(0) \rangle = 2k_B T k i$ to ensure stationarity, otherwise the OU noise will asymptotically thermalize due to the FDT. Each independent mode relaxes exponentially $\langle \zeta_i(t)\zeta_j(t') \rangle = 2k_B T \delta_{ij} k_i e^{-\nu_i|t-t'|}$ and the total colored noise is given by the sum $\xi(t) = \sum_i^N \zeta_i(t)$.. The

corresponding memory kernel decays as a sum of exponentials with varying timescale, $\eta(t) = \sum_i^N k_i e^{-\nu_i t}$ where the ratio $\{k_i/\nu_i\}$ determines the weight of each mode's contribution to the overall friction. When the relaxation timescales are logarithmically spaced as $\nu_i = \nu_0/b^{i-1}$, the sum kernel approximates power-law memory. Here b is the base of the spacing parameter (we use b = 10 for decade scaling) and $\nu_0$ is the high frequency cut-off for $\xi(t)$, typically chosen slightly below $1/\Delta t$ (39). To approximate the viscoelastic memory kernel, $\eta(t) = \frac{\eta_\alpha}{\Gamma(1-\alpha)} \frac{1}{t^\alpha}$, we use OU parameters from Ref (39) with $k_i = C_\alpha(b)\, \eta_\alpha \nu_0^\alpha / [b^{\alpha(i-1)} \Gamma(1-\alpha)]$. The constant $C_\alpha(b)$ is the dimensionless fitting parameter to ensure the correct scaling across logarithmically spaced modes. In practice, we generate a noise trajectory $\lceil 1/\alpha \rceil$ times longer than the desired simulation steps and then downsample by the same factor. This strategy improves the handling of high-frequency components and ensures accurate statistics across time scales. We adopted N=16 OU modes, which has been shown to accurately approximate a power law over 15 decades of time, a strategy well-supported by prior work (26). The generated noise is normalized and scaled as per the FDT to get the desired noise strength $\xi(t)$. Such construction of $1/f^{1-\alpha}$ thermal noise from a sum of OU noise is standard and widely accepted in numerical GLE simulations (18, 26, 39).

**Trajectory Generation:** The single particle trajectories governed by the GLE with power-law frictional kernel include both a deterministic and a stochastic contribution. The deterministic part depends on the initial configuration, $x(0)$, which is modulated over time by the relaxation function, $\theta(t)$, which in our case is the one-parameter Mittag-Leffler function. For stationary initial conditions, position is sampled from a zero mean unbiased Gaussian distribution with the variance set by the trap $\langle x^2(0) \rangle = \frac{k_B T}{\kappa}$. The stochastic part arises from the convolution of thermal noise $\xi(t)$ with the response function $\chi(t)$. We compute the convolution with discrete time-steps:

$$x(t_n) = x(0)\theta(t_n) + \sum_{k=0}^{n-1} \int_{t_k}^{t_{k+1}} \chi(t_n - t')\xi(t')\, dt' \tag{B2}$$

We approximate the noise $\xi(t)$ as a piecewise constant within each subinterval $[t_k, t_{k+1}]$, which gives:

$$x(t_n) \approx x(0)\theta(t_n) + \sum_{k=0}^{n-1} \xi(t_k) \int_{t_k}^{t_{k+1}} \chi(t_n - t')\, dt' . \tag{B3}$$

This is a causal discrete-time convolution of the colored noise with the response function, which is zero for $t < 0$. To evaluate the response contribution in each sub-interval, we rewrite the GLE as a convolution with the derivative of a single function, $\varphi(t)$:

$$x(t) = x(0)\theta(t) + \int_0^t \varphi'(\tau)\xi(t-\tau)\,d\tau \tag{B4}$$

so that $\chi(t) = \varphi'(t)$. In our case, $\chi(t)$ is the two parameter Mittag-Leffler function. Using the results from Gorenflo et al., we use the identity (Eq 4.4.4 in Ref. (47))

$$\int_0^t s^{\alpha-1} E_{\alpha,\alpha}[-as^\alpha]\,ds = t^\alpha E_{\alpha,\alpha+1}[-at^\alpha] \tag{B5}$$

to define $\varphi(t)$:

$$\varphi(t) = \frac{t^\alpha}{\eta_\alpha} E_{\alpha,\alpha+1}\left[-\left(\frac{t}{\tau_D}\right)^\alpha\right] \tag{B6}$$

Finally, we discretize the full solution as:

$$x(t_n) \approx x(0)\theta(t_n) + \sum_{k=0}^{n-1} \xi(t_k)\,[\varphi(t_{k+1}) - \varphi(t_k)] \tag{B7}$$

We numerically solve this form, which has a discrete convolution of the noise with the difference of the integrated kernel, essentially an approximation of the original convolution integral. In practice, we scaled the noise to reproduce a specified $D_\alpha$ for a fixed $\alpha$ in sub-diffusion (without confinement). Confinement was then introduced for this sub-diffusive parameter set $\{D_\alpha, \alpha\}$ to achieve relaxation times comparable to those observed in experiments. Fig. S3 shows the TAMSD computed over 1000 numerically simulated trajectories for both free and confined sub-diffusion. Heterogeneity is incorporated by varying the sub-diffusion parameters $D_\alpha$ from well-defined distributions to match the $P(\Delta x, \varepsilon)$ (see Main Text). All the simulations are performed in MATLAB 2022a including this causal convolution using the *filter* function in a numerically stable manner.

## III. Supplementary Figures:

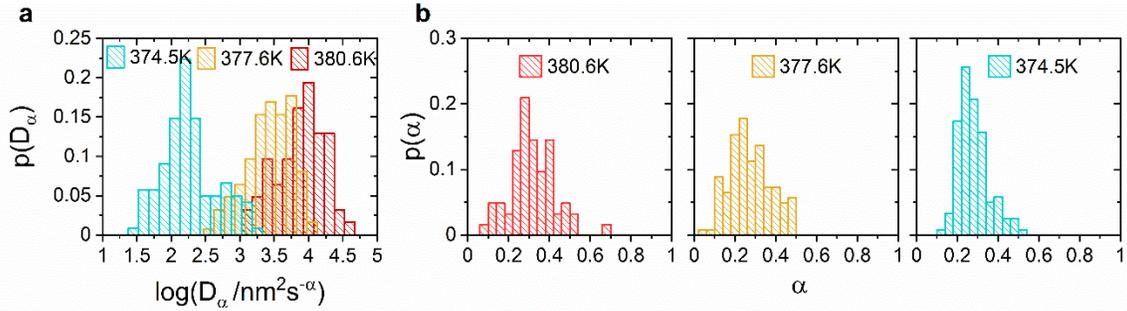

**Fig. S1.** Statistics of sub-diffusion transport parameters, $\{D_\alpha, \alpha\}$, obtained from the initial TAMSD of single molecule trajectories near $T_g$. Shown are the distributions (a) of $\log(D_\alpha)$ and (b) anomalous exponent $\alpha$ at three temperatures: 380.6K (red), 377.6K (yellow) and 374.5K (cyan).

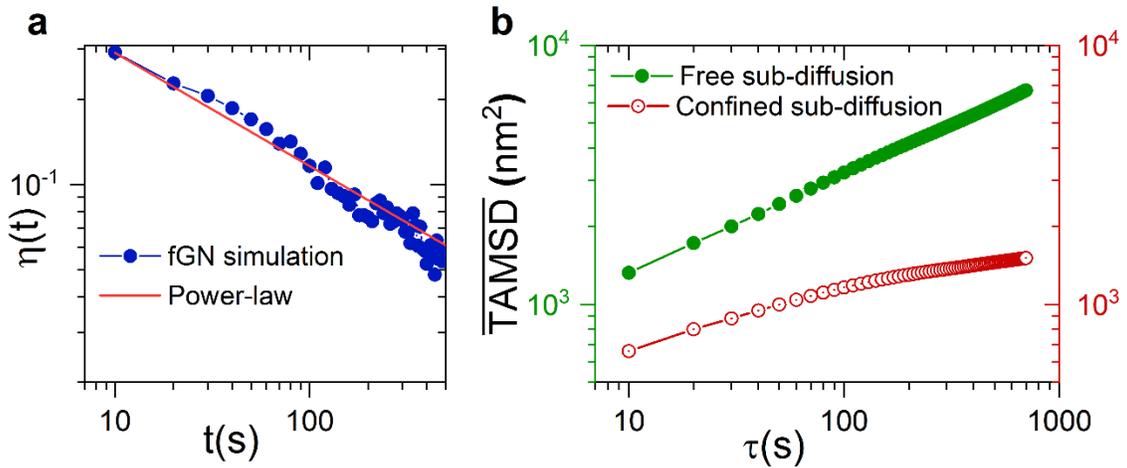

**Fig. S2.** (a) Normalized autocorrelation of fractional Gaussian noise (fGN), obtained from numerical simulation of a power law memory kernel with $\alpha = 0.4$. The solid red line denotes the theoretical power-law memory form. (b) Ensemble-averaged TAMSD from 10000 simulated trajectories for $\alpha = 0.4$, each of length 5000 time-points with $\Delta t = 10s$. Filled green circles correspond to free sub-diffusion, while red circles show confined sub-diffusion with the same $\alpha$ and $\tau_D = 62s$. The comparison highlights the distinct dynamical signatures for free and confined motion under the same memory kernel.

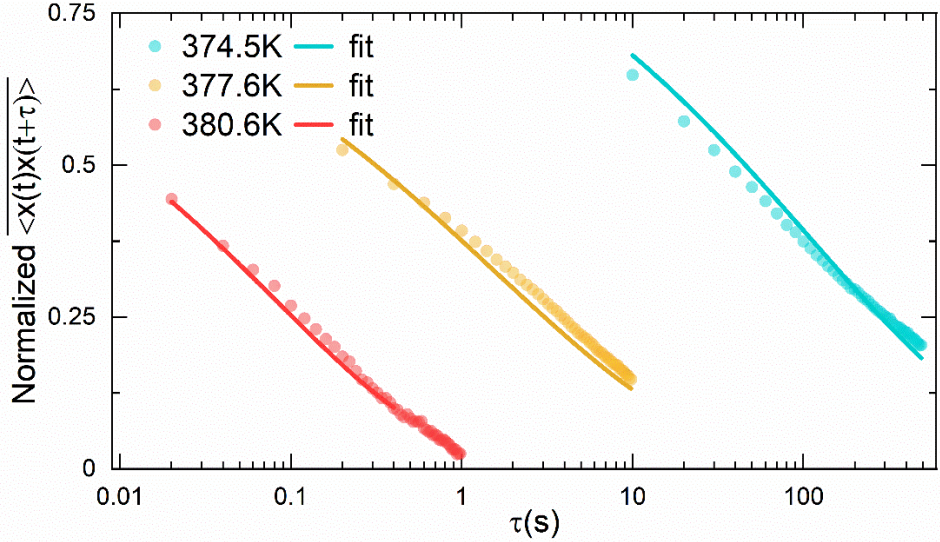

**Fig. S3.** Normalized autocorrelation of position fluctuations near $T_g$ calculated from normalized $\langle x(t)x(t+\tau)\rangle$ averaged over multiple trajectories. Experimental data are shown as filled circles, while solid lines represent the fits obtained with fixed $\alpha$ values determined independently from displacement rate correlation functions. The red, yellow and cyan symbols and lines correspond to measurements at 380.6K, 377.6K and 374.5K, respectively. The first 50 experimental points were used to avoid noise in correlation functions computation at longer time lags.

**Acknowledgments**

This research was supported by National Science Foundation under grant number CHE 2246765. We thank Prof. David R. Reichman for helpful discussions and Prof. Jaesung Yang for reviewing the manuscript.